\def\papertitle{Automatic Equalization for Individual Instrument Tracks Using Convolutional Neural Networks} 
\def\paperauthorA{Florian Mockenhaupt}
\def\paperauthorB{Joscha Simon Rieber}
\def\paperauthorC{Shahan Nercessian}

\documentclass[twoside,a4paper]{article}
\usepackage{etoolbox}
\usepackage{dafx_24}
\usepackage{amsmath,amssymb,amsfonts,amsthm}
\usepackage{euscript}
\usepackage[T1]{fontenc}
\usepackage[utf8]{inputenc}
\usepackage{ifpdf}
\usepackage[english]{babel}
\usepackage{caption}
\usepackage{subfig} 
\usepackage{color}
\usepackage{booktabs}
\usepackage{float}

\input glyphtounicode
\ninept

\newcounter{numauth}
\newcounter{listcnt}
\newcommand\authcnt[1]{\ifdefined#1 \stepcounter{numauth} \fi}

\newcommand\addauth[1]{
\ifdefined#1 
\stepcounter{listcnt}
\ifnum \value{listcnt}<\value{numauth}
\appto\authorslist{, #1}
\else
\appto\authorslist{~and~#1}
\fi
\fi}
\authcnt{\paperauthorB}
\authcnt{\paperauthorC}
\authcnt{\paperauthorD}
\authcnt{\paperauthorE}
\authcnt{\paperauthorF}
\authcnt{\paperauthorG}
\authcnt{\paperauthorH}
\authcnt{\paperauthorI}
\authcnt{\paperauthorJ}
\def\authorslist{\paperauthorA}
\addauth{\paperauthorB}
\addauth{\paperauthorC}
\addauth{\paperauthorD}
\addauth{\paperauthorE}
\addauth{\paperauthorF}
\addauth{\paperauthorG}
\addauth{\paperauthorH}
\addauth{\paperauthorI}
\addauth{\paperauthorJ}

\usepackage{times}

\newif\ifpdf
\ifx\pdfoutput\relax
\else
   \ifcase\pdfoutput
      \pdffalse
   \else
      \pdftrue
\fi

\ifpdf 
  \usepackage[pdftex,
    pdftitle={\papertitle},
    pdfauthor={\authorslist},
    pdfsubject={Proceedings of the 27th International Conference on Digital Audio Effects (DAFx24)},
    colorlinks=false, 
    bookmarksnumbered, 
    pdfstartview=XYZ 
  ]{hyperref}
  \usepackage[pdftex]{graphicx}
\else 
  \usepackage[dvips]{epsfig,graphicx}
  \usepackage[dvips,
    pdftitle={\papertitle},
    pdfauthor={\authorslist},
    pdfsubject={Proceedings of the 27th International Conference on Digital Audio Effects (DAFx24)},
    colorlinks=false, 
    bookmarksnumbered, 
    pdfstartview=XYZ 
  ]{hyperref}
\fi
\usepackage[hypcap=true]{caption}
\title{\papertitle}

\affiliation{
\paperauthorA$^{1}$, \paperauthorB$^{1}$, and \paperauthorC$^{2}$\thanks{\vspace{-3mm}}}
{\href{https://www.native-instruments.com/}{Native Instruments} \\ $^{1}$Langenfeld, Germany $^{2}$Boston, MA, USA \\
{\tt \{\href{mailto:florian.mockenhaupt@native-instruments.com}{florian.mockenhaupt}|\href{mailto:joscha.rieber@native-instruments.com}{joscha.rieber}|\href{mailto:shahan.nercessian@native-instruments.com}{shahan.nercessian}\}@native-instruments.com}
}

\begin{document}
\ifpdf 
  \DeclareGraphicsExtensions{.png,.jpg,.pdf}
\else  
  \DeclareGraphicsExtensions{.eps}
\fi


\maketitle
\begin{abstract}
We propose a novel approach for the automatic equalization of individual musical instrument tracks. Our method begins by identifying the instrument present within a source recording in order to choose its corresponding ideal spectrum as a target. Next, the spectral difference between the recording and the target is calculated, and accordingly, an equalizer matching model is used to predict settings for a parametric equalizer. To this end, we build upon a differentiable parametric equalizer matching neural network, demonstrating improvements relative to previously established state-of-the-art. Unlike past approaches, we show how our system naturally allows real-world audio data to be leveraged during the training of our matching model, effectively generating suitably produced training targets in an automated manner mirroring conditions at inference time. Consequently, we illustrate how fine-tuning our matching model on such examples considerably improves parametric equalizer matching performance in real-world scenarios, decreasing mean absolute error by 24\% relative to methods relying solely on random parameter sampling techniques as a self-supervised learning strategy. We perform listening tests, and demonstrate that our proposed automatic equalization solution subjectively enhances the tonal characteristics for recordings of common instrument types.
\end{abstract}

\section{Introduction}

Equalizers (EQs) form an important class of audio processors capable of directly increasing or decreasing the loudness of specific frequencies within an audio signal \cite{välimäki2016}. The EQ is a fundamental tool for shaping the tone of sound, and apart from leveling and panning, EQ processing is essential to ensuring the harmonious blending of multiple elements within a musical mix \cite{zoelzer2008digital}. One of the most commonly found forms of EQs is the parametric EQ, owing to its high level of user control, low-latency operation, and lack of pre-ringing artifacts \cite{nercessian2020}. A parametric EQ is characterized by multiple controllable EQ bands, each specified by their type (shelving, peaking, high/low-pass filters, etc.), frequency, gain, and quality factor (Q) \cite{abel-berners2004}.

Automatic EQ algorithms determine suitable tonal treatments to be applied to a given audio recording, with the presence or even in lieu of a reference recording characterizing the tonal style of interest \cite{steinmetz2022style}. Such treatments can be achieved by predicting settings for some underlying parametric EQ processor \cite{iriz2023}, or by means of an end-to-end model which directly applies equalization (and potentially other effects \cite{ramirez2022}) to the audio in a non-parametric manner \cite{ramirez2018}. Here, we focus our attention on the former, in order to promote predictable and artifact-free processing, computational efficiency, and seamless integration into existing parametric EQs, offering interpretable control and automation functionalities to end users of varying skill levels. Under this paradigm, a basic automatic EQ objective may be one that aims to balance all frequencies evenly, as in \cite{perez-gonzalez2009}. As demonstrated in \cite{davis2002}, it is evident that different instrument types have their own characteristic spectral properties and therefore, targeting a linear frequency response (e.g. a pink noise target) across all forms of instrument tracks is unlikely to be suitable. Accordingly, more targeted criteria can be derived, potentially by data-driven means \cite{deman2017}. Specifically, the development of an effective automatic EQ system then becomes intimately related to realizing a suitable parametric EQ matching solution, whereby EQ parameters are inferred so that their resulting frequency response most closely resembles some idealized target treatment.

Signal processing-based EQ matching algorithms have been proposed for both graphic \cite{vesa2017} and fully parametric \cite{abel-berners2004} EQs. With the widespread adoption of neural networks (NN) across intelligent music production systems, several deep learning approaches have also been considered. Earlier deep learning-based methods attempted to minimize rudimentary losses defined over predicted and target EQ parameters \cite{valimaki2019neurally, Vesa2, sony2020}. In our own past work \cite{nercessian2020}, we demonstrated that a multi-layer perceptron (MLP) network optimized using differentiable biquadratic filters (biquads) significantly improved matching performance relative to such methods, enabling objective functions to be defined directly in terms of target spectra and the spectra derived from predicted EQ parameters. More recently, DeepAFX-ST \cite{steinmetz2022style} generalized upon this concept even further, proposing a novel method to predict parameters for a processing chain consisting of both a parametric EQ and a compressor in order to mimic the production style of a reference audio clip. This approach leveraged differentiable implementations of both processors in order to optimize an end-to-end training objective defined over time-domain waveforms.

Regardless of the audio processors which they automate, or the respective objective functions which are optimized during their training, a common deficiency to parametric matching algorithms based on deep learning continue to be the self-supervised learning strategy that they employ in order to generate training data, whereby training targets are ultimately created by ascribing random configurations to audio effects. While this minimizes data engineering efforts, it inevitably imposes the need to determine suitable distributions from which to sample random parameters, which more often than not are merely chosen by intuition. Meanwhile, previous deep learning-based filter design studies have outlined the impact of parameter distribution selection on model performance, and accordingly, how they may potentially fail in mismatched environments \cite{iirnet}. Therefore, it is quite likely that resulting systems could struggle to generalize in real-world scenarios, whereby desired target treatments may deviate considerably from those seen during training, reducing their adoption and confidence in their overall effectiveness when deployed in an automated music production context.

In this paper, we propose a system to automatically equalize individual instrument tracks. At inference time, our algorithm identifies the instrument present in an input audio recording using a suitably trained convolutional neural network (CNN), and uses the resulting instrument class prediction to select a spectral target from a bank of instrument-specific spectral distributions that we have aggregated over an annotated set of produced audio samples. Consequently, it is instructive to note that our automatic EQ does not rely on reference audio examples, unlike several existing approaches \cite{steinmetz2022style, sony2020}. Next, we calculate the log-magnitude spectrum of the input audio and subtract it from this ideal spectrum. The resulting difference spectrum describes how frequencies need to be cut or boosted by an underlying parametric EQ in order to match the desired spectral target. A parametric EQ matching model subsequently tries to predict EQ parameters whose resulting frequency response most closely matches the spectral difference curve taken as its input. Lastly, an appropriately configured parametric EQ set to parameter values inferred from the matching model serves as a back end which processes the input audio signal, resulting in the automatically treated audio.

At the core of our proposed system is its parametric EQ matching model. To this end, we illustrate several improvements relative to established state-of-the-art matching techniques, building upon our own previous work which enabled direct optimization of loss functions defined in the spectral domain \cite{nercessian2020}. Accordingly, we introduce an enhanced CNN-based model architecture, reworking its activations and objectives for its training. Most notably however, we suggest a two-stage training procedure, whereby we directly leverage real-world audio data as training examples during a fine-tuning phase, which more closely resemble demonstrative inputs that would be encountered during practical use of our system. In its most basic form, we can derive realistic training targets for our model in an automated fashion by computing spectral differences between labeled training samples and their corresponding ideal instrument target spectra. In doing so, we effectively reduce the domain mismatch observed between training and real-world inference scenarios, whereby our matching model is intrinsically and more effectively optimized for the latter, as compared to existing deep-learning based audio processor automation methods that rely on synthetic data generation techniques to construct their training data.

The remainder of this paper is organized as follows. Section \ref{sec:autoeq} gives a brief overview of our proposed automatic EQ system, Section \ref{sec:training-of-eq-matching-neural-networks} provides an in-depth explanation of our enhanced parametric EQ matching model and its improved training procedure, Section \ref{sec:datasets} describes the datasets used for training parametric EQ matching models in this work, Section \ref{sec:results} evaluates methods by both quantitative and qualitative means, and Section \ref{sec:conclusions} draws conclusions based on these findings.

\section{Automatic equalization system}
\label{sec:autoeq}
\begin{figure}
  \centering
  \includegraphics[width=0.9\columnwidth]{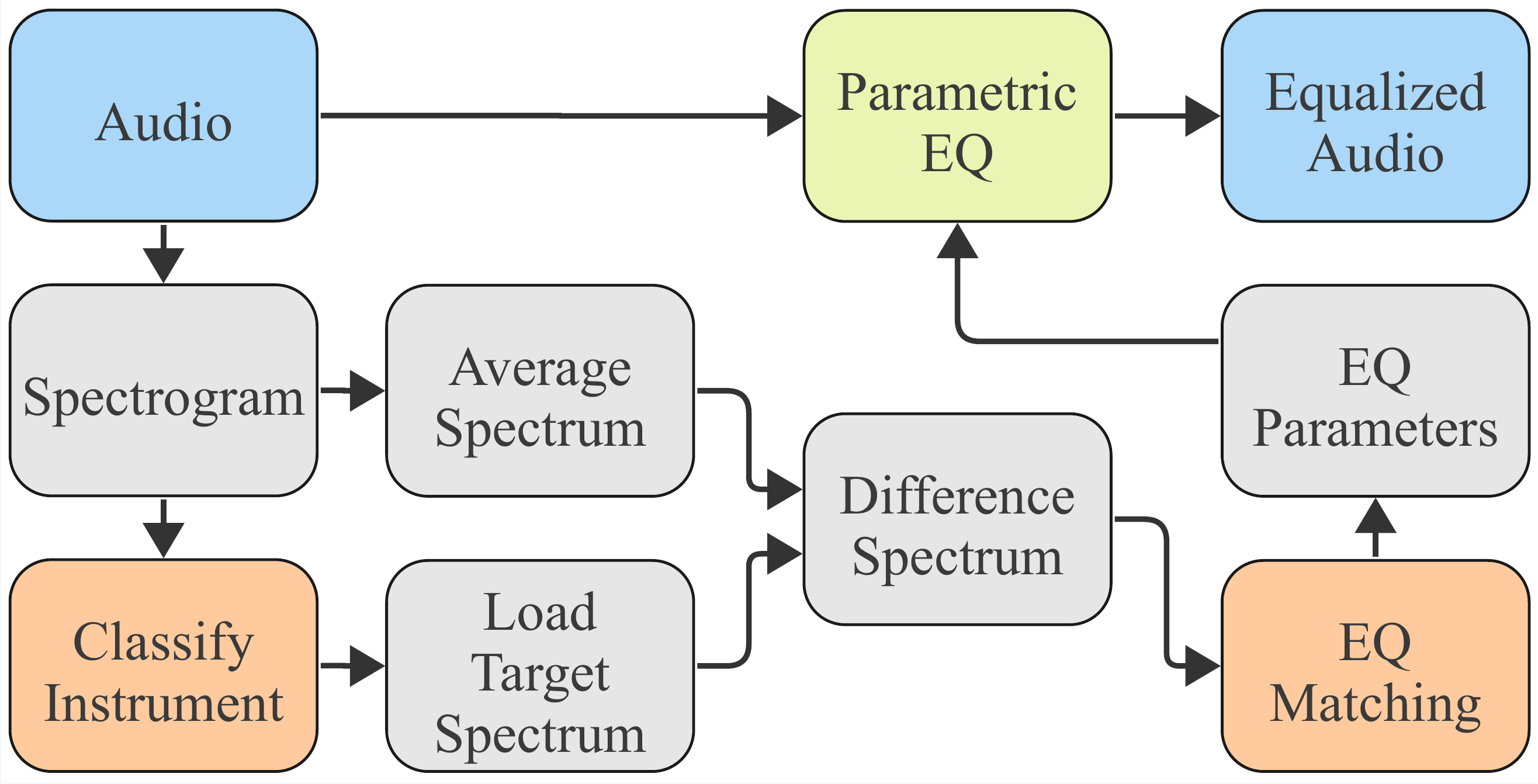}
  \caption{Proposed automatic EQ system.}
  \label{fig:Auto-EQ-system}
\end{figure}

Our automatic EQ method is depicted in Figure \ref{fig:Auto-EQ-system}, consisting of multiple stages in order to predict EQ parameters from a given input audio recording that ultimately drive an underlying parametric EQ processor. We first identify the instrument contained in the input audio recording using a CNN-based instrument classifier model. According to its predicted class, an instrument-specific spectral curve is selected as an ideal target. Next, we compute the spectrum of the input audio and subtract it from the target curve. The resulting curve is then processed by a CNN-based EQ matching model in order to predict suitable EQ parameters. These settings can then be used by an appropriately configured parametric EQ in order to create the automatically equalized audio. In this work, the various analysis blocks within our system assume a nominal audio sample rate $f_\text{s} = 44.1$~kHz, while the parametric EQ processor itself naturally adapts to arbitrary sample rates.

\subsection{Spectrogram Computation}
\label{subsec:Spectrogram-function}
Spectrograms serve as a common input for various components within our system. We compute a spectrogram by first segmenting the signal using Hann windows of size 2048 with an overlap of 50\%. Each window is then processed by computing its discrete Fourier transform and transforming the magnitude spectrum onto the decibel (dB) scale. We then derive a vector $\mathbf{f}$ consisting of 256 logarithmically spaced frequency bins ranging from 20~Hz to 22~kHz, and linearly interpolate the spectrogram onto this grid.

\subsection{Instrument Classifier}
\label{subsec:Instrument-classifier}
In order to classify the instrument type within a source recording, we trained a ResNet-like CNN \cite{resnet2015} with approximately 2M parameters and 35 output instrument classes, including several common instrument types such as electric guitar, piano, vocal, drums and/or various percussive elements. Our internal training dataset covered roughly 45k instrument-annotated audio samples with a sample rate of 44.1 kHz. To prepare our training data for model ingestion, we split audio samples into 6 second segments. We compute the spectrogram for each segment as described in Section \ref{subsec:Spectrogram-function}, resulting in a 256 $\times$ 256 matrix as input for the CNN. We evaluated the classifier on 10k additional test samples that were equally distributed across classes, and achieved a classification accuracy of 79\%. As the majority of incorrect predictions occur between similar classes, e.g. ride cymbals get frequently misclassified as hi-hats, and these instruments tend to have similar spectral properties, such classification errors were not found to be significant for our use case.

\subsection{Instrument-specific Ideal Spectra}
\label{subsec:Instrument-specific-ideal-spectra}
\begin{figure}
  \centering
  \includegraphics[width=\columnwidth]{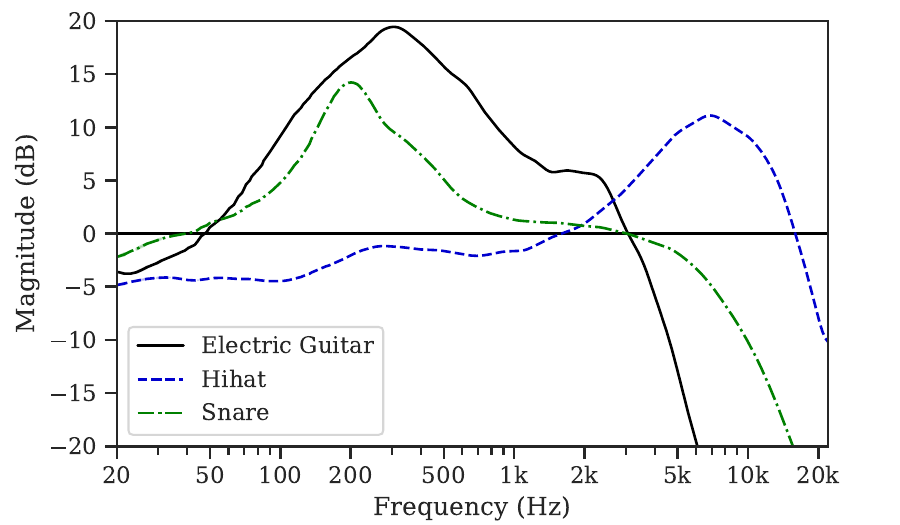}
  \caption{Examples of instrument-specific target spectra derived from our training set.}
  \label{fig:Examples-for-instrument-target-spectra}
\end{figure}

We derive ideal instrument-specific spectral targets, aggregating spectra across a curated subset of our training dataset as a function of instrument class. We calculate spectra according to the procedure described in Section \ref{subsec:Spectrogram-function}, averaging the result along the time axis. Specifically, we calculate the spectrum for each sample, and compute the average spectrum corresponding to each instrument class accordingly. Lastly, we normalize all resulting curves to be zero-mean on the dB scale. Some select spectral target curves derived according to this procedure are illustrated in Figure \ref{fig:Examples-for-instrument-target-spectra}.

\subsection{Spectral Difference Calculation}
\label{subsec:Spectral-difference-calculation}
\begin{figure}
  \centering
  \includegraphics[width=\columnwidth]{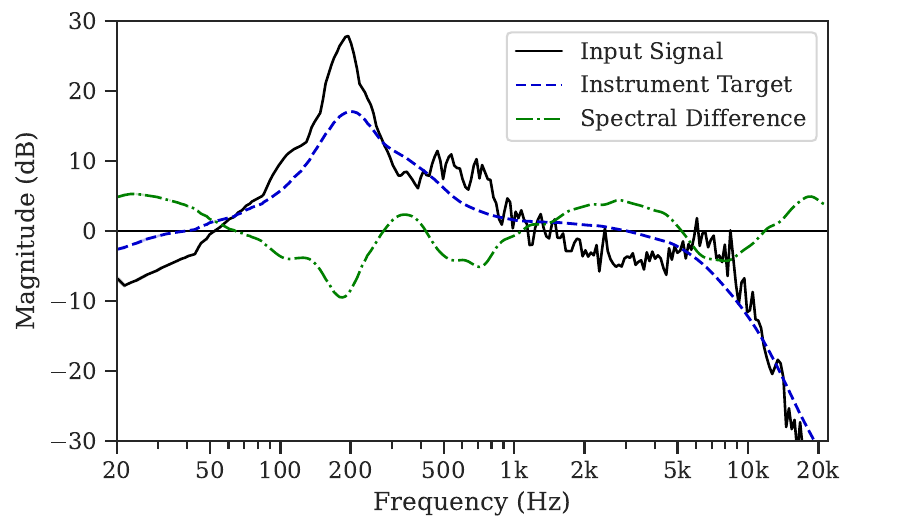}
  \caption{Spectral difference computation for a snare drum input signal with its corresponding ideal instrument target spectrum.}
  \label{fig:Examples for computing the spectral difference}
\end{figure}

We can determine the parts of the frequency spectrum whose loudness should ultimately be boosted or cut by the parametric EQ by performing a spectral difference on the dB scale between the input audio and its corresponding ideal instrument target. We first subtract the measured spectrum of the input audio from the target spectrum. Next, we apply Gaussian smoothing with a kernel standard deviation $\sigma = 3$ in order to filter out local peaks and disregard noisy ripples in the spectral difference. Accordingly, we focus on matching global spectral properties instead of compensating for local resonances. After smoothing, we normalize all resulting curves to be zero-mean on the dB scale, and limit their maximum absolute value to 12 dB by means of scaling. Figure \ref{fig:Examples for computing the spectral difference} illustrates our described procedure for a select input signal/instrument target pair.

\subsection{Parametric EQ Matching Model}
\sloppy We use a parametric EQ matching model to determine a set of EQ parameters whose resulting frequency response closely matches a given input spectral difference curve, as calculated in the previous stage. As such, the output parameters predicted by our model are designed to correspond exactly to the configuration of an underlying parametric EQ processor within our overall system. Since the parametric EQ matching model is such a pivotal component to our overarching method, we dedicate the entirety of Section \ref{sec:training-of-eq-matching-neural-networks} to outline its architecture and training procedure in further detail.

\subsection{Parametric EQ Design}

Parametric EQs are most often implemented as a cascade of biquads, i.e. second-order infinite impulse response filters (IIR) \cite{williston2009digital}. Each biquad implements the difference equation
\begin{align}
\begin{split}
y[n] = {b_0}x[n]+{b_1}x[n-1]+{b_2}x[n-2] \\ - {a_1}y[n-1]-{a_2}y[n-2]
\end{split}
\end{align}
and has the transfer function
\begin{equation}
\label{eqn:transfer}
H(z) = \frac{b_0+b_1z^{-1}+b_2z^{-2}}{1+a_1z^{-1}+a_2z^{-2}},
\end{equation}
where the roots of the numerator and denominator polynomials dictate the nature of its filtering \cite{IzoKorn}. The resulting frequency response of a biquad can be evaluated over a vector of digital frequencies $\mathbf{\omega}$ by evaluating equation (\ref{eqn:transfer}) with $ e^{j\mathbf{\omega}}$ as its argument.

While biquads can model arbitrary second-order digital filters, there exist explicit formulae for defining parametric peaking and shelving filters with specified center/cutoff frequency $f$ in Hz, gain $g$ in dB, and Q/slope $Q$ at a given sampling rate $f_\text{s}$ \cite{RBJ}. To this end, it is common to use the bilinear transform in order to derive digital filters from their analog prototypes. A downside of this is that warping of the response is introduced for center/cutoff frequencies near the Nyquist frequency, as described in \cite{massberg2011digital}. Thus, we use compensated second-order peak and shelving filters as proposed in \cite{orfanidis1997digital} and \cite{abel2003discrete}. These filter designs provide a closer approximation of the analog frequency response, while being more efficient than using oversampling.

A $K$-band parametric EQ can be created by cascading $K$ biquads in series, where each biquad corresponds to an EQ band controlled by the user \cite{christensen2003}. As in previous works \cite{nercessian2020, steinmetz2022style, sony2020}, we consider a parametric EQ with $ K \geq 2$ bands, consisting of exactly one low shelf (band 1), one high shelf (band $K$), and $K-2$ peaking filters (bands 2 through $K-1$). The composite transfer function of the parametric EQ is then given by
\begin{equation}
H_{eq}(z) = \displaystyle \prod_{k=1}^{K} H_{k}(z)
\end{equation}
and the magnitude frequency response of the parametric EQ can be computed as
\begin{equation}
\label{eqn:freqz}
\left|H_{eq}(e^{j\omega})\right| = \left|\prod_{k=1}^{K} H_{k}(e^{j\omega})\right|.
\end{equation}
Here, we evaluate magnitude frequency responses at $\mathbf{\omega} = (\pi/f_s) \mathbf{f}$ and express them on the dB scale, effectively matching the logarithmic frequency grid and output scale used in our various spectral computations.

In this work, we demonstrate our approach with $K=4$, resulting in a parametric EQ comprised of a low shelf, 2 peak filters, and a high shelf. The bands of our parametric EQ have fixed individual parameter ranges, as detailed in Table \ref{tab:filter_ranges}. Gain parameters are to be within the range of $\pm$ 12 dB range across all bands. Frequency ranges are band-specific, with boundary values selected based on their common use in audio mixing contexts. The Qs for shelving filters are set to a fixed value of 0.75, noting that a variety of parametric EQs leverage shelving filters having fixed or default Qs that are at or near this value. The range of Qs for peaking filters spans from 0.1 to 3.0, as in \cite{nercessian2020}. Ultimately, this means that our automatic EQ algorithm is responsible for determining a total of 10 novel EQ parameters, corresponding to 4 frequencies, 4 gains, and 2 Qs. Processing of the input audio signal through a parametric EQ with these resulting settings yields the automatically equalized audio output of our system.
	%
\begin{table}[bt]
\centering
  \caption{\itshape Parametric EQ parameter ranges.}
  \begin{tabular}{|l|c|c|c|c|}
  \hline
      & $\mathrm{Low~Shelf}$ & $\mathrm{Peak~1}$ & $\mathrm{Peak~2}$ & $\mathrm{High~Shelf}$ \\ \hline
      $f_\text{min}$ & 30 Hz & 200 Hz & 600 Hz & 1.5 kHz \\
      $f_\text{max}$ & 450 Hz & 2.5 kHz & 7 kHz & 16 kHz \\
      \hline
      $g_\text{min}$ & -12 dB & -12 dB & -12 dB & -12 dB \\
      $g_\text{max}$ & +12 dB & +12 dB & +12 dB & +12 dB \\ 
      \hline
      $q_\text{min}$ & 0.75$^*$ & 0.1 & 0.1 & 0.75$^*$ \\
      $q_\text{max}$ & 0.75$^*$ & 3.0 & 3.0 &  0.75$^*$ \\ 
      \hline

  \end{tabular}
 \label{tab:filter_ranges}
 \\
$^*$ Fixed/not inferred by our parametric EQ matching model.
\\
\end{table}

\section{Neural Parametric Equalizer Matching}
\label{sec:training-of-eq-matching-neural-networks}
\begin{figure}
  \centering
  \includegraphics[width=\columnwidth]{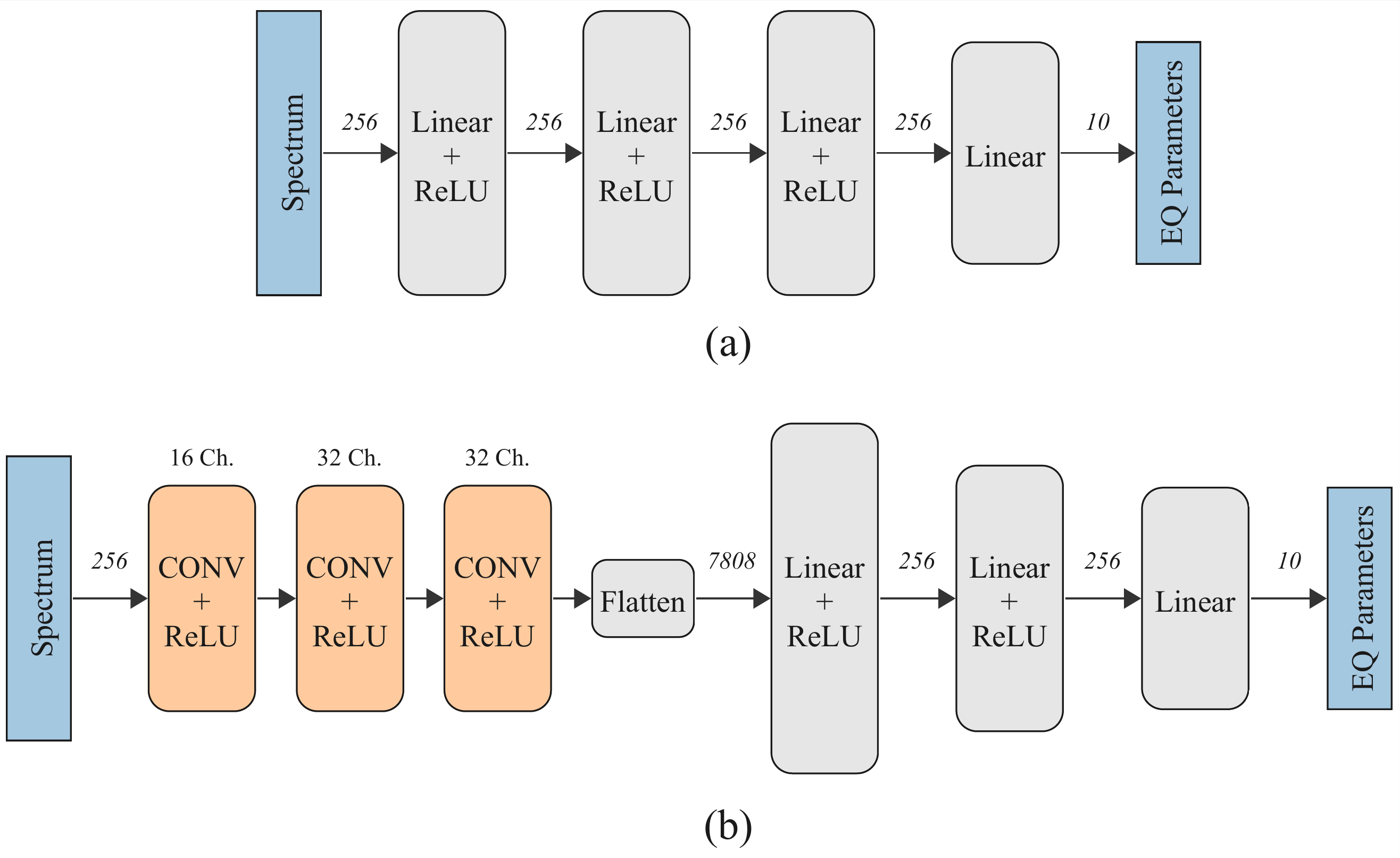}
  \caption{Model architectures for parametric EQ matching:
  (a) baseline MLP model as in \cite{nercessian2020} and 
  (b) our enhanced CNN model with added convolutional front end.}
  \label{fig:Neural Network Architectures}
\end{figure}

\begin{figure}
  \centering
  \includegraphics[width=\columnwidth]{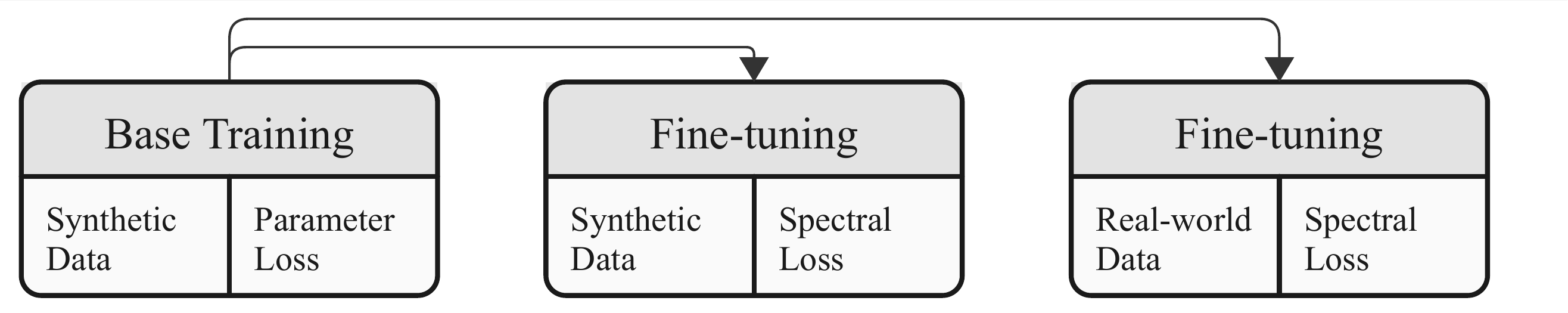}
  \caption{Base training and fine-tuning stages of neural parametric EQ matching models.}
  \label{fig:Training-and-fine-tuning}
\end{figure}

As described in \cite{nercessian2020}, deep learning approaches for parametric EQ matching are capable of outperforming classical approaches \cite{abel-berners2004} under appropriate training conditions. Accordingly, we consider two different NN architectures in this work, as shown in Figure \ref{fig:Neural Network Architectures}. Moreover, we consider a multi-stage training procedure leveraging various combinations of loss terms and data sources, as indicated in Figure \ref{fig:Training-and-fine-tuning}. Base models are first trained on synthetic data using a parameter loss. These base models are then fine-tuned using a spectral loss function, and trained on either synthetically generated or real-world data  for comparison.

\subsection{Model Architectures}
\label{subsec:EQ-matching-model-architectures}
The first model we consider is an MLP similar to the one described in \cite{nercessian2020}, serving as a baseline for this work. Its architecture consists of 3 linear layers with 256 neurons and ReLU activation functions in each layer, which process spectral difference curves as input to the model. The final output linear layer has 256 input and 10 output neurons, corresponding to the variable parameters of our 4-band parametric EQ.

We also propose a second, enhanced model in the form of a CNN, which combines the baseline architecture with an additional convolutional front end. The rationale for such a front end is to learn useful local patterns from desired frequency response curves seen during training. Architecturally, we exchange the first linear layer with 3 convolutional layers with 16, 32 and 32 channels, each followed by a ReLU activation function. All convolutional layers have a kernel size of 5, a stride of 1 and no padding. The number of neurons in the first linear layer is increased to 7808 according to the length of the flattened tensor after the convolutional layers. Subsequent hidden and output layers are identical to the MLP.

Another upgrade that we consider in this work is a deliberate decision to avoid activations at the output of the model entirely, unlike our own past works and other related follow-ons \cite{nercessian2020, steinmetz2022style, IzoKorn}. 
We noted that applying such activations in this context can make models susceptible to vanishing gradients during training, ultimately limiting performance. Naturally, we desire to constrain the values that EQ parameters can take on to some desired range, but we empirically find that we can achieve this more effectively using the combination of techniques outlined in Sections \ref{subsec:eq-parameter-normalization}-\ref{subsec:model-fine-tuning}.  Considering our use of fixed slopes for shelving filters, it turns out that our inferred filter parameterization is guaranteed to yield stable filters without the need for any additional parameter-specific activation functions.  As an illustrative example, we note that inferred frequencies will effectively alias to a frequency within the Nyquist band as a natural consequence of our filter design equations. 

\subsection{EQ Parameter Normalization}
\label{subsec:eq-parameter-normalization}
In order to neutralize the effect that differing value ranges for different EQ parameter types (i.e. frequency, Q, and gain) can have on model training, we normalize all EQ parameters to values between 0 and 1 via
\begin{equation}
	f_{\text{norm}} = \frac{\log\left(\frac{f}{f_\text{min}}\right)}{\log\left(\frac{f_\text{max}}{f_\text{min}}\right)},
\label{eqNormF}
\end{equation}
\begin{equation}
	g_{\text{norm}} = \frac{{g - g_\text{min}}}{{g_\text{max} - g_\text{min}}},
\label{eqNormG}
\end{equation}
and
\begin{equation}
	Q_{\text{norm}} = \frac{{Q - Q_\text{min}}}{{Q_\text{max} - Q_\text{min}}}.
\label{eqNormQ}
\end{equation}
where we use the minimum/maximum parameter boundaries as defined on a band-specific basis for the various EQ parameter types in Table \ref{tab:filter_ranges}. Underlying EQ parameters can be computed from normalized model outputs by applying their respective inverse normalization functions.

\subsection{Base Training Using a Parameter Loss Function}
In a first step, we perform regression with an EQ parameter loss function using synthetically generated spectral difference curves derived from randomized EQ parameter settings (whose sampling strategy will be described in further detail in Section \ref{subsec:Synthetic-data}). This helps the model to predict EQ parameters in the desired normalized range, and provides a starting point for further fine-tuning in a second stage. In this case, the inner workings of the parametric EQ is assumed to be unknown to the model, and as such, there is no direct means of performing backpropogation through the audio processor itself. The parameter loss is defined as
\begin{equation}
    \mathcal{L}_\text{v}(\mathbf{v}, \hat{\mathbf{v}}) = || \mathbf{v}-\hat{\mathbf{v}} || _1,
\label{eqParameterLoss}
\end{equation}
with $\hat{\mathbf{v}}$ and $\mathbf{v}$ as predicted and target EQ parameters, respectively. We posit that model parameters based on this training are certainly better calibrated to the matching task than a random initialization. As noted in \cite{nercessian2020}, optimizing a parameter loss in this manner has been previously considered in related works \cite{valimaki2019neurally, Vesa2, sony2020}.

\subsection{Model Fine-tuning Using a Spectral Loss Function}
\label{subsec:model-fine-tuning}
Motivated by the use of differentiable audio processors to enable end-to-end training \cite{engel2020ddsp}, our previous work in \cite{nercessian2020} already demonstrated that EQ matching performance can be improved by using differentiable biquads to calculate the magnitude frequency response from predicted EQ parameters. The key insight expressed there was that parametric EQ filter design formulae and equation (\ref{eqn:freqz}) (with corresponding $\mathbf{\omega}$) could readily be implemented using differentiable operations within deep learning libraries, allowing for backpropogation through a loss defined directly in terms of output magnitude frequency responses and model input spectral difference curves. Accordingly, we can define a spectral loss as
\begin{equation}
    \mathcal{L}_\text{x}(\mathbf{x}, \hat{\mathbf{x}}) = || \mathbf{x}-\hat{\mathbf{x}} || _1
\label{eqSpectralLoss}
\end{equation}
with $\hat{\mathbf{x}}$ as the resulting frequency response (on the dB scale) calculated from predicted EQ parameters $\hat{\mathbf{v}}$. The spectral difference curve $\mathbf{x}$ naturally serves as both the input to the model as well as a training target (as is the case with an auto-encoder).

\begin{figure}
  \centering
  \includegraphics[width=\columnwidth]{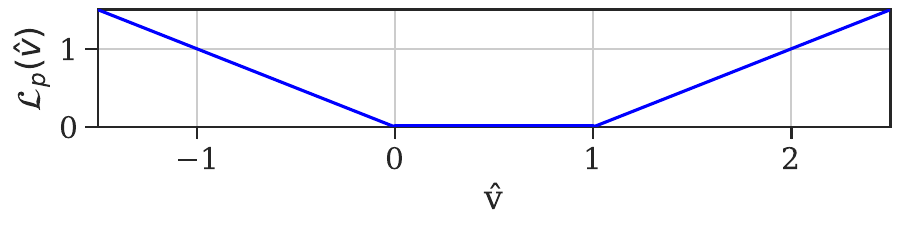}
  \caption{Penalty loss term for fine-tuning.}
  \label{fig:Penalty loss}
\end{figure}

As previously mentioned, the bands of our parametric EQ have fixed individual parameter ranges. Accordingly, we introduce an additional loss term to our training objective during fine-tuning to ensure that the predicted EQ parameters are within these defined ranges, foregoing the need for output activation functions which may contribute to vanishing gradients. This penalty term is shown in Figure \ref{fig:Penalty loss}, and is defined as
\begin{equation}
    \mathcal{L}_\text{p}(\hat{\mathbf{v}}) = ReLU(-\hat{\mathbf{v}}) + ReLU(\hat{\mathbf{v}}-1).
\label{eqPenaltyLoss}
\end{equation}

The net effect of this added term is if a predicted parameter is out of its desired range during training, the absolute difference between the predicted parameter and the exceeded boundary is added to the loss, bearing resemblance to penalty methods used in some constrained optimization algorithms \cite{penalty}. Finally, the composite objective function used during model fine-tuning is
\begin{equation}
    \mathcal{L}_\text{f}(\mathbf{x}, \hat{\mathbf{x}}, \hat{\mathbf{v}}) = \mathcal{L}_\text{x}(\mathbf{x}, \hat{\mathbf{x}}, \hat{\mathbf{v}}) + \lambda\mathcal{L}_\text{p}(\hat{\mathbf{v}}),
\label{eqFinetuningLoss}
\end{equation}
where we set $\lambda = 1$ in this work for demonstrative purposes. We can fine-tune models according to this objective using either synthetically generated data or using spectral difference curves derived from real-world data, and compare both approaches in this work.

\section{Datasets}
\label{sec:datasets}
In this section, we describe the construction of datasets used to train our parametric EQ matching models, as well as to assess both parametric EQ matching performance and the fidelity of equalized audio products from our composite automatic EQ algorithm. In addition to synthetic data generation mechanisms using random parameter sampling techniques as in several previous works \cite{nercessian2020, steinmetz2022style, valimaki2019neurally, sony2020}, we demonstrate how real-world data can also be utilized during parametric EQ matching model training due to the system design of our automatic EQ method.

\subsection{Synthetic Data Via Random EQ Parameter Sampling}
\label{subsec:Synthetic-data}
We leverage synthetically generated data for the base training of the EQ matching model, and as an alternate data source during fine-tuning for comparative purposes. In this case, we generate input spectral difference curves to serve as model input by randomly sampling EQ parameters and compute their corresponding magnitude response. The normalized representation of these EQ parameters can be utilized as training targets in the base training case using a parameter loss function. Their corresponding frequency responses are used as model inputs, and alternatively as training targets in the fine-tuning case making use of a spectral loss function. Random frequencies are determined for each EQ band by
\begin{equation}
    f_{\text{rand}} = f_{\text{min}} \cdot e^{x_{\text{rand}} \cdot \log\left(\frac{f_{\text{max}}}{f_{\text{min}}}\right)}
\label{eq44}
\end{equation}
with $x_{\text{rand}} \sim \mathrm{Uniform}(0, 1)$. This guarantees the same probability for all frequencies on a logarithmic scale. Random EQ band gains are calculated as
\begin{equation}
    g_{\text{rand}} = x_{\text{rand}}^3 \cdot g_{\text{max}} \cdot x_{\text{sign}}
\label{eq6}
\end{equation}
with $x_{\text{rand}} \sim \mathrm{Uniform}(0, 1)$, as well as a random multiplicative sign term $x_{\text{sign}} \sim 2\cdot\mathrm{Bernoulli}(0.5)-1$ which leads to negative gains with a probability of 50\%. The $x_{\text{rand}}^3$ term results in a higher probability for selecting small absolute gain values, as we expect lower gains to occur more frequently in real-world applications. Shelving filters have a fixed Q of 0.75, and for peak filters, we generate random Q values $Q_{\text{rand}} \sim \mathrm{Uniform}(q_\text{min}, q_\text{max})$. Lastly, in order to achieve more natural results for our generated spectra, we perform augmentation by adding random noise to each amplitude value. Using this procedure, we generate a training dataset containing 156,730 synthetically generated curves with their associated EQ parameters.

\subsection{Real-world Data}

While synthetically generated training data can serve as a decent proxy for high-fidelity training examples in closed experimental settings, we would ideally like to train our parametric EQ matching model using real-world data examples resembling those that we are likely to encounter during inference. In order to create such a training dataset, we processed a 31,346 sample subset of our internal dataset, and applied a form of nearest neighbor data augmentation similar to \cite{narita_ganstrument_2023}. Specifically, we measured the spectrum of each audio example, and compute corresponding spectral difference curves, using not only the respective ground truth instrument target spectrum, but those of 4 other closely related target spectra derived from similar instrument classes for each example. The use of "nearby" instrument target curves in this manner creates additional input/target pairs for training that could still conceivably occur in practical applications. Using this augmentation technique, we create 5 spectral difference curves per audio example, resulting in a total number of 156,730 training input spectra. Additionally, we curate a held-out test dataset consisting of 8k samples and their respective spectral difference curves, which we use to evaluate the various models trained in this work.

At inference time, we leverage our instrument classifier to determine the instrument type of input recordings to our system (and therefore their corresponding spectral target) in the absence of a corresponding ground truth label. As long as our instrument classifier is adequately accurate, we can make the claim that in this instance, training examples based on real-world audio data are more likely to resemble those seen during inference time, as compared to training examples based solely on synthetically generated curves via random parameter sampling. Therefore, one would intuit that fine-tuning on real-world audio data in this manner should improve parametric EQ matching performance as it pertains to practical use cases within our automatic EQ system, since we have explicitly sought out to bridge this domain mismatch gap. Moreover, training examples based on real-world audio data are more likely to be indicative of suitably produced tonal treatments of individual instrument tracks, which we assert has good implications with regards to the treated audio that would ultimately be created by our automatic EQ method.

\section{Experimental Results}
\label{sec:results}
We perform a number of experiments in order to demonstrate the effectiveness of both our enhanced parametric EQ matching capabilities as well as our proposed automatic EQ system taken as a whole. Our quantitative evaluation investigates the impact of architecture, loss function, and data type on parametric EQ matching model performance, while our subjective evaluation is used to validate whether or not our automatic EQ system can indeed improve tonal characteristics of input audio recordings in a fully automated fashion. All evaluations are performed using our held-out real-world audio data test set, reflecting realistic use cases for our methods.

\subsection{Objective Evaluation of Parametric EQ Matching Models}
We trained a total of 6 models in order to compare parametric EQ matching accuracy across several factors. First, we trained base models using both MLP and CNN architectures. Next, we performed separate fine-tunings on synthetic and real-world data for each respective base model, as described in Section \ref{sec:training-of-eq-matching-neural-networks}. All training stages across all models used the Adam optimizer with a learning rate of $10^{-4}$ and a batch size of 128. We trained models for 3 epochs during each stage while reducing the learning rate after every epoch by a factor of 0.1. Our objective evaluation methodology involved computing the average spectral loss, as assessed over all 8k examples in our real-world test set. Given its definition and domain on which it operates, we simply refer to this as the mean absolute error (MAE) of a given model.

\begin{figure}
  \centering
  \includegraphics[width=\columnwidth]{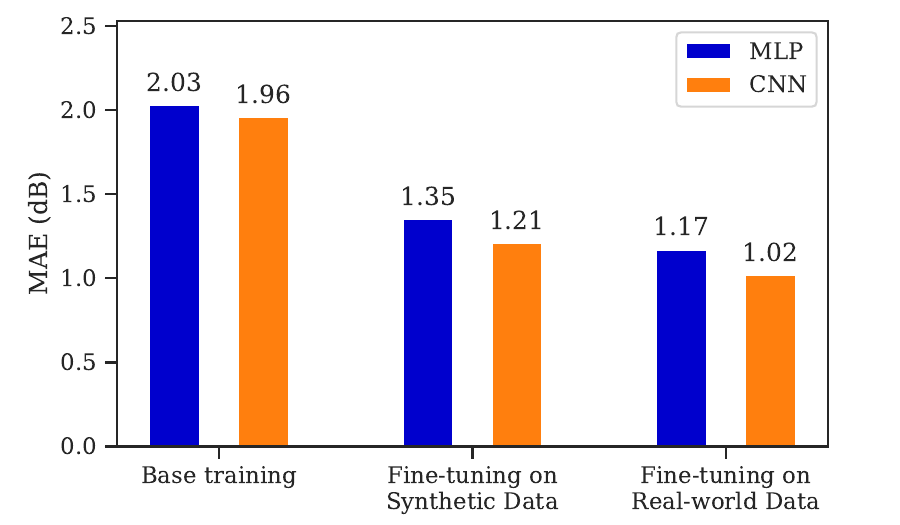}
  \caption{Quantitative evaluation of the neural parametric EQ matching models trained in this work, as measured on our curated test set.}
  \label{fig:Evaluation bar plot}
\end{figure}

Our quantitative evaluation is summarized in Figure \ref{fig:Evaluation bar plot}, revealing several key observations within the different training configurations we tested. While base training with a parameter loss notionally aligns models towards our given task, we observe a significant decrease in MAE when directly optimizing a spectral loss function, even when training on synthetically generated data derived via random EQ parameter settings. Specifically, the reduction amounts to 0.68 dB for the MLP and 0.75 dB for the CNN, aligning with prior findings documented in \cite{nercessian2020}. We note a further decrease in MAE when utilizing real-world training data. The reduction for the MLP and CNN is 0.18 dB and 0.19 dB, respectively. Moreover, the superiority of the CNN architecture over the MLP is evident, with consistent MAE reductions observed across all trainings. Using the CNN, the base training yields an MAE reduction of 0.07 dB against the MLP. When fine-tuning on synthetic or real-world data, the use of the CNN in place of the baseline MLP results in MAE reductions of 0.14 dB and 0.15 dB, respectively.

\begin{figure}
  \centering
  \includegraphics[width=\columnwidth]{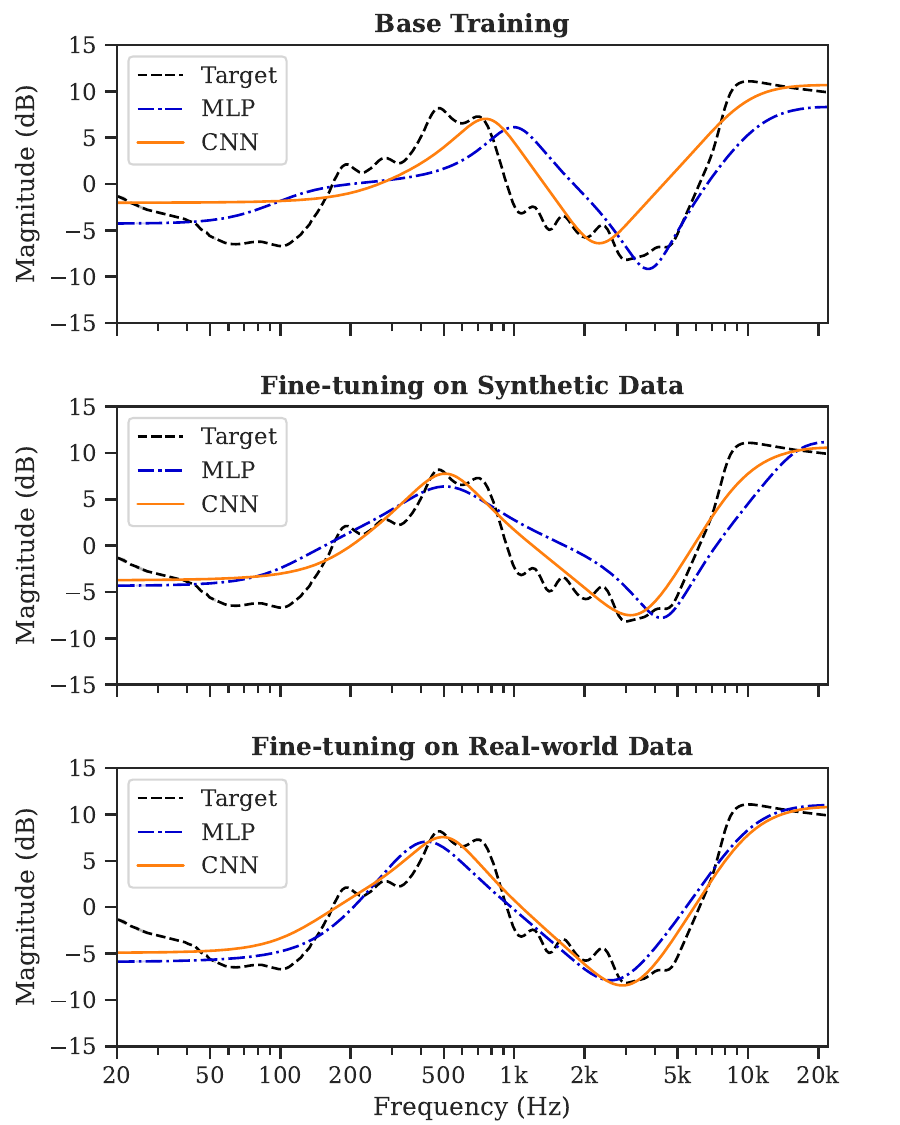}
  \caption{Parametric EQ matching comparisons of the various models trained in this work, demonstrated using a piano recording as input.}
  \label{fig:model_comparision}
\end{figure}

Fine-tuning of the MLP with synthetic data mirrors the approach outlined in \cite{nercessian2020}, recording an MAE of 1.35 dB. Meanwhile, our enhanced CNN architecture yields an MAE of 1.02 dB when fine-tuned on real-world data. Therefore, we exhibit a net decrease of 0.33 dB (a 24\% improvement) in comparison to our previous state-of-the art approach. Figure \ref{fig:model_comparision} illustrates sample outputs for the various models we trained in this work, where we can visually note the improved matching capability of our best performing model over previously published baselines. 

\subsection{Subjective Evaluation of our Automatic EQ System}
We demonstrate the effect of our automatic EQ system on real-world individual instrument recordings by conducting a listening test. This subjective evaluation consisted of 35 test trials from audio examples within our held-out test set, including individual tracks reflecting a number of common instrument types, consisting of a full acoustic drum set, individual drum elements, electric guitar, bass, brass, synths, organ, and vocals. We considered the untreated audio and audio processed with our proposed system (according to predicted EQ parameters from our best performing EQ matching model) as stimuli for each test case. To prevent participants to be biased by different loudness, we normalized all examples to -23 LUFS (loudness units relative to full scale \cite{lufs}). In each trial, participants were tasked to decide whether they prefer the original audio or the automatically equalized one. We encouraged participants to make these decisions based on the following criteria: 1) the individual instrument sounds like it commonly "should," 2) problematic frequencies are suitably suppressed, and 3) the individual instrument track is more likely to fit within a full mix context. Accordingly, we constructed our evaluation as a blind A/B test, whereby participants do not know which audio example is the original one and which one is treated across all test cases, whilst providing a third neutral option to indicate no preference. This is based on the recent findings presented in \cite{camp2023mos}, where the authors compared A/B and mean opinion score listening test formats exemplary of those used to evaluate text-to-speech systems, and showed that A/B tests are more reliable and less influenced by the number of listeners. 

\begin{figure}
  \centering
  \includegraphics[width=0.9\columnwidth]{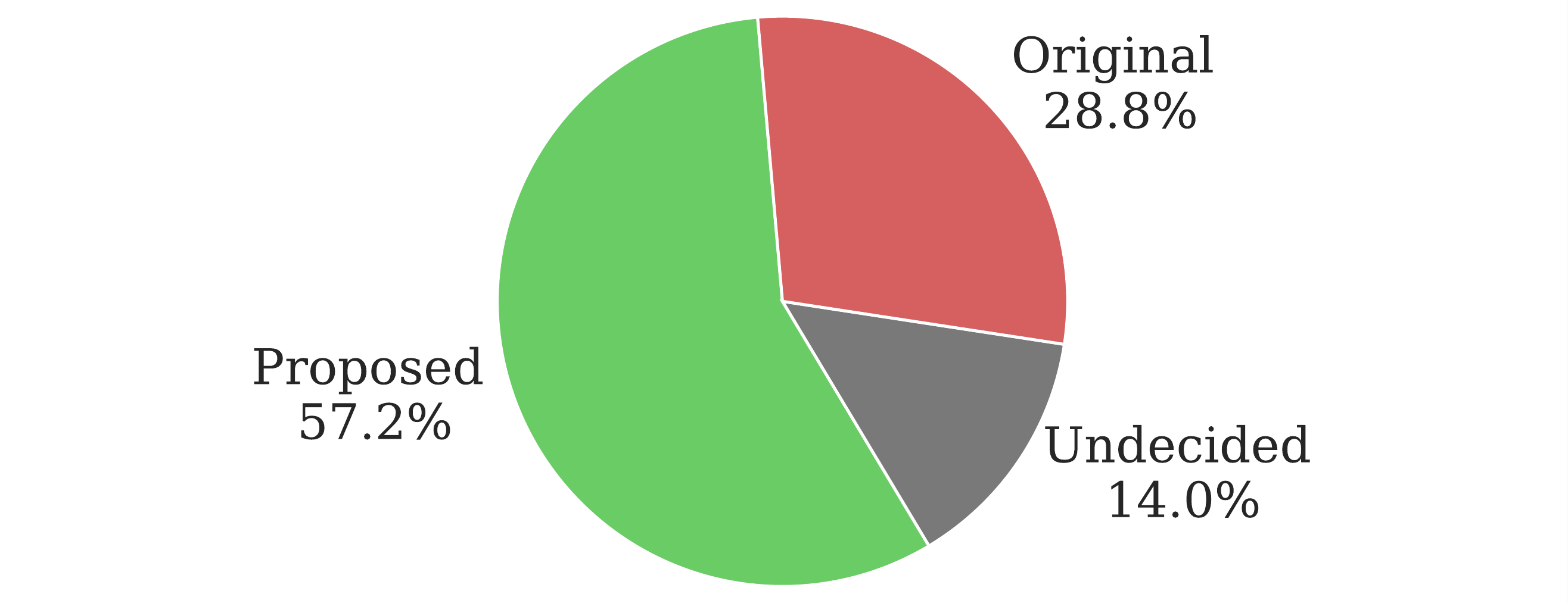}
  \caption{High-level summary of our subjective evaluation.}
  \label{fig:abx}
\end{figure}

The results of our listening tests are summarized in Figure \ref{fig:abx}. In total, 26 participants within our organization (with proficient listening skills and varied musical experience) took part in the evaluation. It is observed that participants preferred the equalized treatments produced by our proposed automatic EQ technique nearly 2 to 1 over the original audio, and were undecided in 14\% of cases.

\section{Conclusions}
\label{sec:conclusions}

In this paper, we proposed an automatic EQ system that predicted settings for an underlying parametric EQ, using a combination of CNNs for parametric EQ matching and instrument classification. We quantitatively showed how our proposed enhancements to neural parametric EQ matching led to fidelity gains in accuracy relative to our previous state-of-the-art approach in \cite{nercessian2020}. This was accomplished by utilizing a CNN architecture, revamping model activations and learning objectives, and lastly, by leveraging real-world audio samples as part of a two-stage training process. To this end, the use of our instrument classifier within the proposed system allowed us to significantly reduce domain mismatch gaps between training and real-world inference scenarios relative to previous methods \cite{steinmetz2022style, sony2020}, increasing the overall effectiveness of our composite system. Accordingly, we conducted listening tests and showed that our proposed automatic EQ system was capable of subjectively enhancing the sound quality for common instrument and vocal recordings, all without the need for providing reference audio at inference time. Future work will be dedicated to additional refinements to our system, and investigating further how our parametric EQ matching paradigm scales to various EQ configurations. We are also interested in devising similar automation systems for other forms of audio processors, such as the direct dynamics processor in \cite{nercessian2022}.


\nocite{*}
\bibliographystyle{IEEEbib}
\bibliography{DAFx24_tmpl} 

\end{document}